\documentstyle[preprint,aps]{revtex}
\begin{document}
\title{{\bf {\it Direct Correlation between Tc and CuO}}$_{2}${\bf {\it \ Bilayer
Spacing in YBa}}$_{2}${\bf {\it Cu}}$_{3}${\bf {\it O}}$_{7-x}$}
\author{M. Varela$^{a,b}$, D. Arias$^{a,\S }$, Z. Sefrioui$^{a}$, C. Le\'{o}n$^{a}$,
C. Ballesteros$^{b}$, S. J. Pennycook$^{c}$ and J. Santamaria$^{a}$}
\address{$^{a}$ GFMC, Dpto Fisica Aplicada III, Facultad de Fisica, Universidad\\
Complutense, Ciudad Universitaria. 28040 Madrid. Spain\\
$^{b}${\it \ Dpto Fisica, Universidad Carlos III de Madrid, Avda de la}\\
Universidad 30, 28911 Legan\'{e}s, Madrid. Spain\\
$^{c}${\it \ Solid State Division, Oak Ridge National Laboratory, Tenessee,}%
\\
U.S.A.\\
$^{\S }${\it On leave from Universidad del Quindio. Armenia. Colombia}}
\date{June 13$^{th}$, 2002}
\maketitle

\begin{abstract}
We report the effects of epitaxial strain and deoxygenation on high quality
[YBa$_{2}$Cu$_{3}$O$_{7-x}$(YBCO) $_{N}$ / PrBa$_{2}$Cu$_{3}$O$_{7}$ (PBCO) $%
_{5}$]$_{1000\text{\AA }}$ superlattices, with 1 
%TCIMACRO{\TEXTsymbol{<}}%
%BeginExpansion
\mbox{$<$}%
%EndExpansion
{\it N}%
%TCIMACRO{\TEXTsymbol{<}}%
%BeginExpansion
\mbox{$<$}%
%EndExpansion
12 unit cells. High spatial resolution electron energy loss spectroscopy
(EELS) shows that strained, fully oxygenated YBCO layers are underdoped.
Irrespective of whether underdoping is induced by strain or deoxygenation,
X-ray diffraction analysis shows that Tc correlates directly with the
separation of the CuO$_{2}$ bilayers.
\end{abstract}

\newpage

Since the discovery of the high Tc superconductivity, the relation between
critical temperature and structure has been the focus of considerable
interest. Changes in the critical temperature (Tc) induced by pressure
(physical) and isovalent substitution (chemical pressure) have been
extensively investigated to find which interatomic distances within the unit
cell are relevant for the mechanism of superconductivity (see for instance
references 1-3 and references therein). However, despite a considerable
experimental and theoretical effort, a definitive answer has not yet been
obtained, partly due to the complex intracell structure of the cuprates.
Most theories consider superconductivity primarily located on the CuO$_{2}$
planes, which are coupled by proximity or pair tunneling effects [4-9], and 
extensive experimental work has been done, directed to examine the
importance of the distance between CuO$_{{\bf 2}}$\ planes of the same and
of neighboring cells for the mechanism of superconductivity.  Modifications
of the interlayer coupling have been explored changing the number of copper
planes [10], their inner structure [11,12], or intercalating non
superconducting spacers as iodized compounds [13] or organic spacers
[14].Concerning the separation between neighboring CuO$_{{\bf 2}}$\ planes,
substitution of the rare earth element is known to influence lattice
dimensions (chemical pressure experiments) [1, 15]. An increase of the rare
earth ionic radius from 0.99 A for Yb to 1.08 for Nd is known to increase
the distance between planes and also Tc from 88 to 95 K [16]. Despite the
extensive work, the results of some of the experiments are
contradictory[13,14], a clear picture has not emerged yet. Changes in the
distance between planes, or modifications in the intracell distances due to
cationic substitution, may have, in addition, a direct influence in the
charge transfer and thus on the carrier concentration{\bf .}On the other
hand it is well known that changes in oxygen content, which result in
changes in the carrier concentration, also cause significant structural
changes [17,18]. This interplay between structural changes and doping has
obscured the relationship between superconductivity and structure.

In addition to physical or chemical pressure, structural changes induced by
epitaxial strain in thin films can be used to explore the relationship
between structure and superconductivity. In a previous paper we have
reported that significant epitaxial strain occurs in ultrathin YBa$_{2}$Cu$%
_{3}$O$_{7-x}$ layers in [YBa$_{2}$Cu$_{3}$O$_{7-x}$(YBCO) $_{N}$ / PrBa$%
_{2} $Cu$_{3}$O$_{7}$ (PBCO) $_{5}$] $_{1000\text{\AA }}$ superlattices,
with N ranging between 1 and 12 unit cells, as a result of a 1\% in-plane
lattice mismatch with PBCO. This epitaxial strain causes non-uniform changes
in the interatomic distances within the unit cell which correlate with the
decrease of the critical temperature in the strained layers [19,20].

In this paper we compare structural changes resulting from epitaxial strain
with those arising from doping in oxygen depleted YBCO samples. In both
cases, changes in the intracell distances are comparable in magnitude but
some have opposite signs, and the comparison provides important information
on the interplay between structure and superconducting properties. Using
high spatial resolution electron energy loss spectroscopy (EELS) in a
scanning transmission electron microscope (STEM) we show, for the first
time, that epitaxially strained layers are underdoped. Structure refinement
by X-ray diffraction shows that regardless of whether underdoping is caused
by strain or deoxygenation, Tc is uniquely determined by the CuO$_{2}$
bilayer separation. No other intracell spacings show such a correlation,
indicating that the CuO$_{2}$ bilayer spacing is the fundamental structural
signature of Tc in the superconducting cuprates.

The samples for this study were epitaxially strained [YBa$_{2}$Cu$_{3}$O$%
_{7-x}$ /PrBa$_{2}$Cu$_{3}$O$_{7}$]$_{1000\text{\AA }}$superlattices grown
by high pressure (3.4 mbar) oxygen sputtering on (100) SrTiO$_{3}$
substrates, held at 900%
%TCIMACRO{\UNICODE{0xba}}%
%BeginExpansion
${{}^o}$%
%EndExpansion
C. The PBCO layer thickness was fixed in 60 \AA , (5 unit cells), enough to
decouple the superconducting layers, while the YBCO layer thickness ranged
between 12 and 1 unit cells. The number of bilayers grown in the stacking
was adjusted to obtain samples 1000\AA\ thick. Structure was analyzed by
x-ray diffraction (XRD) and refined with the SUPREX software [21,22], which
allows refinement of atomic positions in the c- direction within the unit
cell. Electron energy loss spectroscopy (EELS) measurements and Z-contrast
images were obtained in a VG Microscopes HB501UX STEM operated at 100kV with
a field emission gun, and a parallel detection EELS post-column
spectrometer. An anular dark field detector was used for imaging. In this
microscope, the electron probe can be focused down to a diameter of a 2.2\AA
. Cross section samples for STEM were prepared by conventional grinding,
dimpling and ion milling with Ar ions with an energy of 5 kV, at an
incidence angle of 7%
%TCIMACRO{\UNICODE{0xba}}%
%BeginExpansion
${{}^o}$%
%EndExpansion
. Final cleaning was done at low voltage of 2 kV.

Superlattices grown by this technique had atomically flat interfaces with
negligible interdiffusion as proven by x-ray diffraction analysis and energy
filtered transmission electron microscopy. Details about sample preparation
and interface characterization can be found elsewhere [20,23]. In these YBCO
layers, when the thickness is reduced below 4 unit cells epitaxial strain
effects arise due to the 1\% lattice mismatch between YBCO and PBCO, causing
a quite inhomogeneous reorganization of intracell distances [19]. For the
thinnest (one unit cell) YBCO layer, the CuO$_{2}$ planes move closer to
each other by 4\%, while the Ba ion approaches the CuO chains by 4\% but
moves away from the CuO$_{2}$ planes by 2\%. Results for several sets of
samples are shown in Fig. 1. Tc is plotted versus the distance between the
CuO$_{2}$ planes and the Y cation, i.e. half of the distance between
adjacent CuO$_{2}$ planes (fig 1a), and versus the distance between the Ba
ion and the CuO chains (fig 1b). Note that despite the strain within the
single layers might be inhomogeneous, x-ray analysis supplies values for the
intracell distances averaged over the whole layer. The [YBCO$_{N}$/PBCO$_{5}$%
] superlattice samples, with 1%
%TCIMACRO{\TEXTsymbol{<}}%
%BeginExpansion
\mbox{$<$}%
%EndExpansion
N%
%TCIMACRO{\TEXTsymbol{<}}%
%BeginExpansion
\mbox{$<$}%
%EndExpansion
12 unit cells, where the parameter tuning the Tc is epitaxial strain, are
represented by solid squares. The c-lattice parameter of the PBCO was
constant at a value of 11.71 $\stackrel{o}{{\bf A}}$, very close to the bulk
value, pointing to an unstrained structure of the PBCO [19]. Lattice
distortion due to the lattice mismatch with the STO in the first PBCO layer
( +0.4 \% along a axis and -0.6 \% along b axis) was not detected.

It is interesting to compare these structural changes with those arising
through deoxygenation of bulk YBCO by Jorgensen et al [17]. Upon oxygen
depletion the hole density decreases gradually, and so does the value of Tc.
At the same time, quite substantial structural changes occur within the bulk
YBCO unit cell. Contrary to the strained samples, one observes that as
oxygen is removed, the Ba ion moves {\it away} from the basal CuO chain (5\%
for an oxygen content of 6.4 per formula unit), while the CuO$_{2}$ planes 
{\it approach} each other (3\%). Meanwhile, the distance between the Ba ion
and the CuO$_{2}$ planes decreases slightly. Figure 1 demonstrates a clear
correlation between structural changes and variations in the critical
temperature for both deoxygenation and epitaxial strain. It is remarkable
that changes in the different intracell distances caused by strain and
deoxygenation are quantitatively similar, although in the case of the Ba-CuO
distance show opposite sign.

For the superlattices, increasing epitaxial strain decreases the distance
between CuO$_{2}$ planes, as would be expected from the Poisson effect
alone. This is true both for adjacent planes (same cell) and for planes of
neighboring cells [19,20]. It would be expected that the enhanced coupling
would increase Tc, which is contrary to the results observed. This suggests
that a different effect is ruling the changes in the critical temperature.
Significantly, the spacing of the CuO$_{2}$ bilayers {\it decreases} {\it %
also upon oxygen removal}, despite the fact that the c lattice parameter
increases in this case.

The behavior of the Ba-CuO spacing between the Ba atom and the CuO chains
(figure 1b) is more complicated. In epitaxially strained samples Tc
decreases with decreasing Ba-CuO separation, while the opposite trend is
found for the oxygen depleted bulk samples. The bell shaped aspect of figure
1b could point at first glance to an optimal Ba-CuO distance for which the
maximum Tc is attained. Since on the right branch, the oxygen depleted
samples, Tc decreases due to underdoping, the left branch, would be
interpreted as overdoping of the CuO$_{2}$ planes due to the strain induced
structural change. However, this conclusion is opposite to that indicated by
the contraction of the CuO$_{2}$ bilayer spacing which occurs both for
strained and for underdoped samples. In the following we explicitly
deoxygenate the strained superlattices to show that overdoping is {\it not}
the cause of the reduced Tc.

Strained [YBCO$_{1}$/PBCO$_{5}$]$_{1000\text{\AA }}$ superlattices were
depleted of oxygen to a nominal content 7-x= 6.93, 6.8 and 6.6 per formula
unit. Oxygen content was adjusted {\it in-situ}, after sample growth,
following a stability line in the phase diagram corresponding to the desired
oxygen stoichiometry during sample cool down [24]. These oxygen contents are
nominal and the values were inferred from the results obtained on companion
thin films of YBCO which were given the same deoxygenation sequence [25].
The structure was quantitatively determined through the refinement of the
x-ray diffraction spectra. When oxygen was removed, the superlattice Bragg
peaks shifted to lower angles, denoting a monotonic increase of the c
lattice parameter. Although a direct measure of the oxygen content in our
superlattices is not possible, the relative change in the c lattice
parameter obtained from x ray analysis of the superlattices was in perfect
agreement with the values found for bulk samples with the same nominal
compositions.

Figure 2 shows changes of various intracell distances in the superlattices
versus the nominal oxygen content. Results (obtained from x-ray refinement)
are shown for the distance from the CuO$_{2}$ planes to the Y atom (Y-CuO$%
_{2}$), from the CuO$_{2}$ planes to the Ba ion (CuO$_{2}$-Ba), and from the
Ba ion to the chains (Ba-CuO). Deoxygenated superlattices (solid symbols)
follow precisely the same trend as neutron data for bulk samples (open
symbols) from reference [17]. This shows that the structural changes caused
by deoxygenation in strained layers track exactly those occuring in
(relaxed) bulk samples. We can therefore directly test the correlation of
the Ba-CuO spacing to Tc. For a nominal oxygen content of 7-x = 6.6, we have
Tc = 10K in the strained superlattice (see figure 3, discussed below) and
the Ba-CuO spacing attains a value of 2.14\AA . This is close to the value
found for optimally doped {\it unstrained} YBCO with Tc = 90 K (see figure
1b). It can be concluded that Tc is {\it not} determined by the absolute
value of the Ba-CuO spacing.

We now present definitive evidence that the strained superlattices are
underdoped. Figure 3 shows the resistance curves corresponding to the
deoxygenated [YBCO$_{1}$/PBCO$_{5}$]$_{1000\text{\AA }}$ superlattices. The
dotted line corresponds to the fully oxygenated [YBCO$_{1}$/PBCO$_{5}$]
superlattice, and the dashed lines correspond to 7-x = 6.8 and 7-x = 6.6 
{\it in-situ} deoxygenated samples. Slight deoxygenation was also conducted
by {\it ex-situ} annealing the fully oxygenated [YBCO$_{1}$/PBCO$_{5}$]
superlattices at 125%
%TCIMACRO{\UNICODE{0xba}}%
%BeginExpansion
${{}^o}$%
%EndExpansion
C in N$_{2}$ atmosphere for succesive half hour intervals (solid lines in
figure 3) [25]. XRD spectra corresponding to these samples did not show
measurable structural changes. When decreasing oxygen content we can observe
how resistance increases while the Tc decreases smoothly. The reduction in
Tc upon deoxygenation clearly shows that the samples are not overdoped in
their strained, fully oxygenated state.

This result is confirmed using high spatial resolution EELS. Fine structure
on the oxygen K-edge provides a sensitive measure of the superconducting
carrier concentration [26-29]. It is known that the Cu-3d and O-2p states of
the CuO$_{2}$ plane atoms lie close to the Fermi level. As a result of the
dipole selection rule, the unoccupied part of these states can be
investigated by exciting transitions from Cu-2p and O-1s levels - in other
words, by examining the fine structure of the copper L$_{23}$ and oxygen K
edges. In fully oxygenated YBCO two pre-edge features develop at the O-K
edge: a shoulder around 535 eV, which represents transitions to unoccupied
Cu-3d states, and a peak around 528 eV representing transitions to O-2p
states which give rise to holes in the valence band [26-29]. In samples
where the oxygen content is decreased this last peak falls in intensity,
indicating a decrease in hole concentration and a reduction in the critical
temperature. In figure 4 we show the energy loss spectra corresponding to
the O-K edge for two different superlattices: a relaxed [YBCO$_{8}$/PBCO$_{5}
$] sample, with a Tc of 88K (dotted line) and a strained [YBCO$_{1}$/PBCO$%
_{5}$] sample, showing a Tc of 35K (solid line). Spectra were acquired
placing the electron probe on the center of the YBCO unit cell. The inset of
figure 4 displays a Z-contrast image of a [YBCO$_{1}$/PBCO$_{5}$] sample
showing very good resolution of the one unit cell thick layers. While the
pre-peak at 528 eV is present for the relaxed sample, it is absent in the
ultrathin strained layers. This evidences an important change in the hole
density in the YBCO layers in the strained superlattice, confirming that the
hole concentration is substantially reduced in epitaxially strained samples.
Therefore, we can rule out the possibility of overdoping due to strain as a
cause of Tc depression in [YBCO$_{1}$/PBCO$_{5}$] superlattices. It is
possible that the strained layers are underdoped not due to the strain but
as a result of oxygen deficiency provided that the determination of absolute
oxygen concentration in such thin films is a difficult task. However,
annealings in pure oxygen at 1 atmosphere and 500 
%TCIMACRO{\UNICODE{0xba}}%
%BeginExpansion
${{}^o}$%
%EndExpansion
C did not produce significant T$_{{\bf c}}$\ changes. Moreover, these fully
oxygenated samples did not show measureable persisting photoconductivity
after long term illumination with white light [30], thus excluding the
possibility of oxygen deficiency.

Having established that both epitaxially strain and deoxygenation cause
underdoping, we return to the correlation of the CuO$_{2}$ bilayer
separation with Tc shown in Fig. 1a. For our samples this correlation is 
{\it independent} of how a particular spacing was achieved, epitaxial strain
or deoxygenation (or a combination). This shows that {\it the primary
structural signature of doping level and Tc is the CuO}$_{2}${\it \ bilayer
spacing}. This can be interpreted naively as follows: as oxygen is removed
from the chains, electrons are returned to the CuO$_{2}$ planes, which are
therefore more strongly attracted by the trivalent rare earth ion. In
underdoped samples the separation of the CuO$_{2}$ planes is thus
reduced.Our results do not rule out the posibility of other effects such as
fluctuations or size effects being involved in the Tc depression in
ultrathin films [31-35] . But we report new experimental results showing the
connection between Tc and structural modifications.

In summary, we have examined the interplay between epitaxial strain and
doping in strained [YBa$_{2}$Cu$_{3}$O$_{7-x}$(YBCO) $_{N}$ / PrBa$_{2}$Cu$%
_{3}$O$_{7}$ (PBCO)$_{5}$] $_{1000\text{\AA }}$ superlattices. Through
resistivity measurements and high spatial resolution EELS we have shown
directly that strained ultrathin layers are underdoped. Deoxygenation of
strained samples causes non-uniform changes in the intracell structure,
following trends quantitatively and qualitatively similar to those observed
in deoxygenated bulk YBCO. In all cases Tc is directly related to the
separation of the CuO$_{2}$ bilayers. In contrast, the distance between the
Ba atom and the basal plane behaves quite differently with epitaxial strain
and doping and can therefore be used to monitor the interplay between them.
These results should stimulate future theoretical studies to highlight the
role played by the Ba atom position on the charge transfer mechanism.

\newpage {\bf REFERENCES}

\bigskip

[1] A. A. R. Fernandes, J. Santamaria, S. L. Bud%
%TCIMACRO{\UNICODE{0xb4}}%
%BeginExpansion
\'{}%
%EndExpansion
ko, O. Nakamura, J. Guimpel, I. K. Schuller, Phys. Rev. B {\bf 44}, 7601
(1991)

\bigskip

[2] J. P. Attfield, A. L. Kharlanov, J. A. McAllister, Nature {\bf 394}, 157
(1998)

\bigskip

[3] W.E. Pickett. Phys. Rev. Lett. {\bf 78,} 1960 (1997)

\bigskip

[4] S. Chakravarty, A. Sudb\o , P.W. Anderson, S. Strong, Science {\bf 261},
337 (1993),

\bigskip

[5] S. Chakravarty, A. Sudb\o , P.W. Anderson, S. Strong Phys. Rev. B {\bf %
49,} 12245 (1994)

\bigskip

[6] S. Chakravarty and P.W. Anderson, Phys. Rev. Lett. {\bf 72,} 3859 (1994)

\bigskip

[7] P.W. Anderson, Science, {\bf 279,} 1196 (1998)

\bigskip

[8] A.J. Legget Science, {\bf 274}, 587 (1996)

\bigskip

[9] J.L. Tallon, G. V. M. Williams, C. Bernhard, D. M. Pooke, M.P. Staines,
J.D. Johnson, R.H. Meinhold. Phys. Rev. B {\bf 53}, R11972 (1996)

\bigskip

[10] Xiaojia Chen, Chande Gong. Phys. Rev. B {\bf 59} 4513 (1999)

\bigskip

\bigskip \lbrack 11] O. Chmaissem, J.D. Jorgensen, S. Short, A. Knizhnik, Y.
Eckstein, H. Shaked. Nature, {\bf 397} 45 (1999)

[12] K.A. Lokshin, D.A. Pavlov, S.N. Putilin, E.V. Antipov, D.V. Sheptyakov,
A.M. Balagurov. Phys. Rev. B {\bf 63} 064511 (2001)

\bigskip

\bigskip \lbrack 13] X.D. Xiang, W.A. Vareka, A. Zettl, J.L. Corkill, M.L.
Cohen, N. Kijima, R. Gronsky. Phys. Rev. Lett. {\bf 68} 530 (1992)

\bigskip

[14] J.-H. Choy, S.-J. Kwon, G.-S. Park. Science {\bf 280} 1589 (1998)

\bigskip

[15] X.J. Chen, C.D. Gong, Y.B. Yu. Phys. Rev. B {\bf 61}, 3691 (2000)

\bigskip

[16] J.G. Lin, C.Y. Huang, Y.Y. Xue, C.W. Chu, X.W. Cao, J.C. Ho. Phys. Rev.
B {\bf 51} R12900 (1995)

[17] J.D. Jorgensen, S. Pei, P. Lightfoot, D. G. Hinks, B. W. Hinks, B. W.
Veal, B. Dabrowski, A.P. Paulikas, R. Kleb, Physica {\bf C 171,} 93 (1990)

\bigskip

[18] C.C. Almasan, S:H. Han, B.W. Lee, L.M. Paulius, M.B. Maple, B.W. Veal,
J.W. Donwey, A.P. Paulikas, Z. Fisk, J.E. Schirber, Phys.Rev. Lett. {\bf 69,}
680 (1992)

\bigskip

[19] M Varela, Z. Sefrioui, D. Arias, M.A. Navacerrada, M.A. L\'{o}pez de la
Torre, M. Lucia, C. Leon, G.D Loos, F Sanchez-Quesada, J Santamaria. Phys.
Rev. Lett. {\bf 83,} 3936 (1999)

\bigskip

[20] M. Varela, D. Arias, Z. Sefrioui, C. Le\'{o}n, C. Ballesteros, J.
Santamar\'{i}a. Phys. Rev. B {\bf 62,} 12509 (2000)

\bigskip

[21] E. E. Fullerton, I. K. Schuller, H. Vanderstraeten, Y. Bruynseraede,
Phys. Rev. B {\bf 45}, 9292 (1992)

\bigskip

[22] E.E. Fullerton, J. Guimpel, O. Nakamura, I.K. Schuller, Phys.Rev. Lett. 
{\bf 69,} 2589 (1992).

\bigskip

[23] M. Varela, W. Grogger, D. Arias, Z. Sefrioui, C. Le\'{o}n, C.
Ballesteros, K. M. Krishnan, and J. Santamar\'{i}a. Phys. Rev. Lett. {\bf 86}%
, 5156 (2001)

\bigskip

[24] Z. Sefrioui, D. Arias, M. Varela, J.E Villegas, M.A. L\'{o}pez de la
Torre, C. Le\'{o}n, G. D. Loos, J. Santamar\'{i}a.Phys. Rev. B {\bf 60,}
15423 (1999)

\bigskip

[25] It is important to note that while these soft anneals do not cause
deoxygenation in single YBCO films, slight deoxygenation is favored in
superlattices by the strained structure of the ab plane.

\bigskip

[26] N. N\"{u}cker, J. Fink, J.C. Fuggle, P.J. Durha, W. M. Temmerman. Phys.
Rev. B {\bf 37,} 5158 (1988)

\bigskip

[27] N.D. Browning, J. Yuan, and L.M. Brown, 1992. {\it Physica C} {\bf 202}%
, 12 (1992)

\bigskip

[28] N.D. Browning, M.F. Chisholm, D.P. Norton, D.H. Downdes, and S.J.
Pennycook, {\it Physica C} {\bf 212}, 185 (1993)

\bigskip

[29] N.D. Browning, J.P. Buban, P.D. Nellist, D.P. Norton, and S.J.
Pennycook. {\it Physica C} {\bf 294}, 183 (1998)

\bigskip

[30] G. Nieva, E. Osquiguil, J. Guimpel, M. Maenhoudt, B. Wuyts, Y.
Bruynseraede, M. B. Maple and I. K. Schuller, Appl. Phys. Lett. {\bf 60}
2159 (1992).

\bigskip 

[31] C. Bandte. Phys. Rev. B {\bf 49} 9064 (1994)

\bigskip

[32] M. Rasolt, T. Edis, Z. Tesanovic. Phys. Rev. Lett.{\bf \ 66} 2927 (1991)

\bigskip

[33] Y. Matsuda, S. Komiyama, T. Terashima, K. Shimura. Y. Bando. Phys. Rev.
Lett. {\bf 69} 3228 (1992)

\bigskip

[34] T. Terashima, K. Shimura. Y. Bando, Y. Matsuda, A. Fujiyama, S.
Komiyama. Phys. Rev. Lett. {\bf 67} 1362 (1991)

\bigskip

[35] M.Z. Cieplak, S. Guha, S. Vadlamannati, T. Giebultowicz, P. Lindenfeld.
Phys. Rev. B {\bf 50} 12876 (1994)

\newpage {\bf FIGURE CAPTIONS}

\bigskip

Figure 1: (a) Tc of YBCO versus the CuO$_{2}$ bilayer separation, i.e., half
the distance between adjacent CuO$_{2}$ planes (b) Tc versus the distance
between the Ba ion and the CuO chains. Open circles are neutron diffraction
data for deoxygenated bulk YBCO from reference [17]. Black squares represent
the [YBCO$_{N}$/PBCO$_{5}$] superlattices with 12%
%TCIMACRO{\TEXTsymbol{>}}%
%BeginExpansion
\mbox{$>$}%
%EndExpansion
N%
%TCIMACRO{\TEXTsymbol{>}}%
%BeginExpansion
\mbox{$>$}%
%EndExpansion
1 unit cells. Open squares correspond to deoxygenated [YBCO$_{1}$/PBCO$_{5}$%
] superlattices. Dotted lines are a guide to the eye. Vertical dotted lines
correspond to spacings in fully oxygenated bulk samples. Error bars denote
the range within the XRD refinement was not sensitive to changes in the
distances. 

\bigskip

Figure 2: Changes (d-d$_{0}$) of various intracell distances versus the
nominal oxygen content, where d$_{0}$corresponds to the fully oxygenated
bulk samples. Superlattice results (obtained from x ray refinement) are
shown for the Y-CuO$_{2}$ spacing (solid triangles), the CuO$_{2}$-Ba
spacing (solid squares), and the Ba-CuO spacing (solid circles). Neutron
diffraction data for bulk deoxygenated YBCO are taken from reference [17]
(same notation but open symbols).

\bigskip

Figure 3: Resistance versus temperature curves corresponding to the set of
deoxygenated [YBCO$_{1}$/PBCO$_{5}$] strained superlattices. The dotted line
corresponds to a fully oxygenated strained sample. Dashed lines correspond
to nominal oxygen contents of 7-x = 6.8 and 7-x = 6.6. Solid lines represent
the set of slightly deoxygenated superlattices obtained by ex situ
annealings at 125 
%TCIMACRO{\UNICODE{0xba}}%
%BeginExpansion
${{}^o}$%
%EndExpansion
C.

\bigskip

Figure 4: Electron energy loss spectra corresponding to the O-K edge for two
different superlattices: a relaxed [YBCO$_{8}$/PBCO$_{5}$] sample (dotted
line) and a strained [YBCO$_{1}$/PBCO$_{5}$] sample (solid line). The
vertical line shows the position of the pre-peak in the near edge structure
at 528 eV. This peak indicates transitions to O-2p states which give rise to
holes in the valence band. Inset: Z-contrast image corresponding to a [YBCO$%
_{1}$/PBCO$_{5}$] superlattice. The darker contrast corresponds to YBCO
layers (marked with black arrows).

\end{document}